\documentclass[twocolumn,seceq]{jpsj2}

\makeatletter
\title{Large Deviation Property of Free Energy in $p$-Body Sherrington-Kirkpatrick Model}

\author{Tetsuya Nakajima\thanks{tetsuya@huku.c.u-tokyo.ac.jp} and
Koji Hukushima\thanks{hukusima@phys.c.u-tokyo.ac.jp}}

\inst{Department of Basic Science, University of Tokyo, 3-8-1 Komaba, Meguro-ku, Tokyo 153-8902, Japan}
\recdate{\today}
\abst{ 
The cumulant generating function $\phi(n)$ and rate function $\Sigma(f)$ of
free energy is evaluated in the $p$-body Sherrington-Kirkpatrick model
by the replica method with the finite replica number $n$. 
From a perturbational argument, 
we show that the cumulant generating function is constant in the
vicinity of $n=0$. 
On the other hand, with the help of two analytic properties of
$\phi(n)$, the behavior of $\phi(n)$ is derived again. 
However, this is also shown to be broken at a finite value of $n$, which gives a characteristic value in the rate
function near the thermodynamic value of the free energy. 
Through the continuation of $\phi(n)$ as a function of $n$,  we find
a way to derive the 1RSB solution at least in this model, which is to fix the RS solution to
be a monotonically increasing function. 
 }
\kword{$p$-body Sherrington-Kirkpatrick model, large deviation property, replica method}

\begin{document}
\maketitle
\newpage
\section{Introduction}
The Replica Method (RM) is a tool to describe the statics of disordered
 systems. It has been applied mainly to the mean-field models and
 succeeded to provide many interesting concepts. However, one has no
 idea why RM itself is a correct procedure in a mathematical 
 sense and  a satiscaftory answer to the problem
  has not yet been obtained. There are such problems as the choice of saddle point
 or the so-called ``analytic continuation''. These problems are particularly
 difficult when you have to regard them as rigorous. Thus far, only the
 solvable models are studied\cite{OK} or the
 sufficient conditions of 
 the uniqueness of the analytic continuation are examined\cite{vHP}. 
On the other hand, the necessary condition for the continuation has not
 been extensively considered yet, except for ref.~\citen{K}. Thus, as a
 first step toward a better understanding of the RM, we discuss these
 problems through the $p$-body Sherrington-Kirkpatrick model\cite{G}.

Hereafter, we restrict our attention to the systems that have quenched
 disorder and do not discuss the models that do not have 
 explicit randomness, like structural glass. 
For a given disordered system with a Hamiltonian $H$ at the inverse
 temperature $\beta$, the formula of RM is written as 
\begin{equation}
 [\log Z] = \lim_{n\to 0}\frac{1}{n}\log[Z^n], 
\end{equation}
where $Z ={\rm Tr} \exp(-\beta H)$ and  the bracket $[\cdots]$ denotes the
 configurational average with respect to the randomness in the
 system. The moment $[Z^n]$ is evaluated relatively easily when $n$
 is an integer.  
Then, the problem of RM is regarded as the continuation of the cumulant generating function $\phi(n,N) 
:= \frac{1}{n}\log[Z^n]$ from the integer $n$ greater than 1 to
 the vicinity of 0. Here, $N$ denotes the system size.

If the continuation for any real value of $n$ is performed, all the
 information on the probabilistic fluctuation of $\log Z$ is
 obtained\cite{CPSV}. For instance, the variance of the free energy is
 derived as  
\[
\frac{\rm d}{{\rm d} n}\left.\left(\frac{1}{n}\log [Z^n]\right)\right|_{n=0} = [(\log Z)^2]-[\log Z]^2,
\]
and higher cumulants are also derived using corresponding higher order
derivatives. 
Note that an analytic property of $\phi(n)$ in the vicinity of $n=0$ is
significant to the probabilistic fluctuation of the free energy. In
fact, this leads to the large deviation property of the free energy.

Assuming that the self-averaging property for the free energy
 is held, the probability distribution function 
 of $\log Z$ in the limit $N \to\infty$ is expressed as 
 \begin{equation}
 \mathrm{Pr}(\log Z \approx Nf) \approx \exp(-N\Sigma (f)), 
 \label{pdf}
 \end{equation}
where $\Sigma (f)$ is determined using the Legendre transformation, 
 \begin{equation}
 \Sigma (f) = nf - n\phi(n), \hspace{12pt} f = \frac{{\rm d}(
  n\phi(n))}{{\rm d} n}, 
 \label{Legendre}
 \end{equation}
with $n\phi(n) := \lim_{N\to\infty}\frac{1}{N}n\phi(n,N) =  \lim_{N\to\infty}\frac{1}{N}\log [Z^n]$. 
 This is because the original large deviation property describes the probability distribution function of
 the statistical average asymptotically:
 \begin{eqnarray*}
 \mathrm{Pr}\left(\sum_{i=1}^M\log Z^{(i)} \approx MNf\right) \approx \exp(-MN\Sigma (f))  \hspace{12pt}
 \\\mbox{as}
\hspace{12pt} M,N \to\infty, 
 \end{eqnarray*}
where $Z^{(i)}$ is the i.i.d. partition function. 
By setting $f=\frac{1}{N}\log Z$, the formula to determine
 $\Sigma(f)$ is derived by the saddle point method as 
 \begin{eqnarray*}
 \log [Z^n] &=&\log \int{\rm d} fP(f) \exp[nNf]\\
&=&\log \int {\rm d} f \exp\left[N\left\{nf -\Sigma(f)\right\}\right] \\
	&\approx& N\max_f(nf -\Sigma(f)).
 \end{eqnarray*}

The large deviation property of the free energy is evaluated in
 ref.~\citen{PR} for the two-body Sherrington-Kirkpatrick model. In the
 present paper, we apply the framework of ref.~\citen{PR} to the case of
 $p$-body interaction and see the similarity to
 ref.~\citen{MR}, in which the number of metastable states in  the 
 $p$-body SK model is evaluated. They are conceptually different but we
 have found that the way to study the 
 large deviation property proposed in ref.~\citen{PR} is the same as that to calculate the complexity proposed in 
 ref.~\citen{M} for some cases.

The organization of this paper is as follows. In \S\ref{sec:model},
we introduce the model and formulate a calculus of  the cumulant generating function by RM with the
replica number $n$ finite. In \S\ref{sec:perturbation}, following
ref.~\citen{PR}, we  perform the perturbation for the cumulant
generating function with respect to $n$ around $n=0$ where the first-step
replica symmetry breaking (1RSB) ansatz is correct. In \S\ref{sec:connection}, we discuss the continuation between the RS
solution for large $n$ and the 1RSB solution for small $n$. 
In \S\ref{sec:rate}, the rate function $\Sigma$ is
explicitly calculated.  
Finally, \S\ref{sec:summary} is devoted to our conclusions. 

\section{Our Model and Replica Analyses } 
\label{sec:model}

The model Hamiltonian we discuss is given as
\begin{equation}
H = -\frac{1}{\sqrt{N^{p-1}}} \sum_{i_1<\cdots <i_p}J_{i_1\cdots i_p}
S_{i_1}\cdots S_{i_p},  
\end{equation}
where $S_i = \pm 1$ and $J_{i_1\cdots i_p}$ is an
independent random variable following the Gaussian distribution
$N(0,\frac{p!}{2})$. We focus on the case $p\geq 3$. By calculating $[Z^n]$, we can write $\phi(n)$ as\cite{N}
\begin{eqnarray}
\phi(n)=&\frac{\beta^2}{2n}\sum_{\alpha<\beta}^n q_{\alpha\beta}^p -\frac{1}{n}\sum_{\alpha<\beta}^n
q_{\alpha\beta}\hat q_{\alpha\beta} + \frac{\beta^2}{4}  \nonumber\\
&+\frac{1}{n}\log {\rm Tr} \exp\left[ \sum_{\alpha<\beta}^n\hat
q_{\alpha\beta}S^\alpha S^\beta\right], 
\end{eqnarray}
where $q_{\alpha\beta}$ and $\hat q_{\alpha\beta}$ are determined by the
saddle point equations of $\phi$. By imposing the replica symmetric (RS) ansatz,
the solution is obtained as 
\begin{eqnarray}
\phi_0(n;q,\hat q) &= \frac{\beta^2}{4}(n-1)q^p - \frac{n-1}{2}q\hat q + \frac{\beta^2}{4} + \log 2 \nonumber\\
&-\frac{\hat q}{2}+\frac{1}{n}\log\int{\rm D} u\cosh^n(\sqrt{\hat q}u)
\label{phiRS}
\end{eqnarray}
with the saddle point equations 
\begin{eqnarray*}
q &=& \frac{\int{\rm D} u \cosh^n(\sqrt{\hat q}u)\tanh^2(\sqrt{\hat
q}u)}{\int{\rm D} u \cosh^n(\sqrt{\hat q}u)}, \def\nonumber{\global\@eqnswtrue}         \\
\hat q &=& \frac{\beta^2}{2}pq^{p-1},  \def\nonumber{\global\@eqnswtrue}
\end{eqnarray*}
where ${\rm D} u$ represents the normalized Gaussian integral. The RS
solution is considered to be valid for a larger value of $n$ and higher
$T$. 

On the other hand,  based on the 1RSB ansatz, the 1RSB solution is given as 
\begin{eqnarray*}
\phi_1(n;q_0,\hat q_0,q_1,\hat q_1,m) = \frac{\beta^2}{4}\left\{(n-m)q_0^p + (m-1)q_1^p\right\} \\
- \frac{1}{2}\left\{(n-m)q_0\hat q_0+(m-1)q_1\hat q_1\right\}
 +\frac{\beta^2}{4} + \log 2 -\frac{\hat q_1}{2}\\
+ \frac{1}{n}\log\int{\rm D} u\left\{\int{\rm D} v\cosh^m(\sqrt{\hat
			      q_0}u +\sqrt{\hat q_1-\hat
			      q_0}v)\right\}^{n/m}. \def\nonumber{\global\@eqnswtrue}
         \label{phi1RSB}
\end{eqnarray*}
To write the saddle point equations concisely, we define some averages as
\begin{eqnarray*}
\langle f\rangle_1 &:=& \frac{\int{\rm D} v \cosh^m\Xi f(v)}{\int{\rm D} v
\cosh^m\Xi}, \\
\langle f\rangle_0 &:=& \frac{\int{\rm D} u\left\{ \int{\rm D}
v\cosh^m\Xi\right\}^{n/m}f(u)}{\int{\rm D} u\left\{ \int{\rm D}
v\cosh^m\Xi\right\}^{n/m}},  
\end{eqnarray*}
with $\Xi = \sqrt{\hat q_0}u+\sqrt{\hat q_1 - \hat q_0} v$. Then, the
saddle point equations can be expressed as 
\begin{eqnarray*}
q_0 &=& \langle\langle\tanh\Xi\rangle_1^2\rangle_0\hspace{12pt},\hspace{12pt} \hat q_0 =  \frac{\beta^2}{2}pq_0^{p-1}\label{q0det}\def\nonumber{\global\@eqnswtrue}        \\
q_1 &=& \langle\langle\tanh^2\Xi\rangle_1\rangle_0\hspace{12pt},\hspace{12pt} \hat q_1 =  \frac{\beta^2}{2}pq_1^{p-1}\label{q1det}\def\nonumber{\global\@eqnswtrue}        
\end{eqnarray*}
\begin{equation}
\frac{\beta^2}{4}(p-1)(q_1^p-q_0^p)m = \langle\langle\log\cosh\Xi\rangle_1\rangle_0 -\frac{1}{m}\langle\log\int{\rm D} v \cosh^m\Xi\rangle_0.\def\nonumber{\global\@eqnswtrue}
\end{equation}

\section{Perturbation with Respect to $n$}
\label{sec:perturbation}
According to ref.~\citen{G}, the 1RSB solution is valid for $n = 0$ when the
temperature is $T_G<T<T_c$, where $T_c$ is the
paramagnetic-spin glass transition temperature and $T_G$ is the Gardner 
transition temperature\cite{G}, which corresponds to 1RSB-full RSB
transition temperature.  
 This indicates that we have this relation  $ [\log Z] = \lim_{n\to 0}
 \phi_1(n)$. Then, we assume that the 
 1RSB solution 
 describes the behavior of the cumulant generating function in the vicinity of $n=0$ and perform the
 perturbational analysis for the saddle point equations and the cumulant
 generating function itself.

First, we consider the saddle point equations of $q_0(n)$ and $\hat
 q_0(n)$(eq.~(\ref{q0det})). The previous work\cite{G} shows 
 $q_0(0) = \hat q_0(0) = 0$, so we expand them as
 \begin{eqnarray*}
 q_0(n) &=& a_1 n + a_2 n^2 + \cdots \\
 \hat q_0(n)&=&\hat a_1 n +\hat a_2 n^2 + \cdots \def\nonumber{\global\@eqnswtrue}
\label{eqn:perturbation}
\end{eqnarray*}
and determine their coefficients from its lower order of $n$. 
Expanding r.h.s. of the former equation of eq.~(\ref{q0det}) by
$\sqrt{\hat{q_0}}$,   we see that its $\hat{q}_0$ dependence is expressed
only as a function of $\sqrt{\hat{q_0}}u$ in the Gaussian integral. The
leading term of the expansion of $q_0$ is found to start from the linear
$\hat{q}_0$ term by performing the integral. 
Then, the coefficient is $a_1 = \alpha \hat a_1$ because there are no
singularities with respect to the other order parameters or the saddle point equation itself. 
By contrast, the latter
equation of eq.~(\ref{q0det}) yields clearly $\hat{a}_1=0$ because
$p-1\geq 2$. It turns out that $a_1=\hat{a}_1=0$ and this leads again to
$\hat{a}_2=a_2=0$ in a similar manner. 
Thus, we conclude inductively that $a_k=\hat a_k = 0$ for all $ k$ and $q_0(n) = \hat q_0(n) \equiv 0$.

Substituting the result for $q_0$ and $\hat q_0$ in the $\phi_1(n)$, we
 have 
 \begin{eqnarray}
 &\hspace{-30pt}\phi_1(n;0,0,q_1,\hat q_1,m) = \frac{\beta^2}{4}(m-1)q_1^p -
 \frac{1}{2}(m-1)q_1\hat q_1\nonumber\\
&+\frac{\beta^2}{4} + \log 2 -\frac{\hat q_1}{2}+ \frac{1}{m}\log\int{\rm D} v\cosh^m(\sqrt{\hat q_1}v). 
 \label{reduct1RSB}
 \end{eqnarray}
Note that this expression is independent of $n$. Because the order
 parameters $q_1$ and $\hat{q}_1$ are determined by minimizing $\phi_1$,
 they also lose the dependence on $n$. Consequently, we conclude that
 $\phi(n) = \phi(0)$ in the vicinity of $n=0$.

\begin{figure}
\includegraphics[width=8cm]{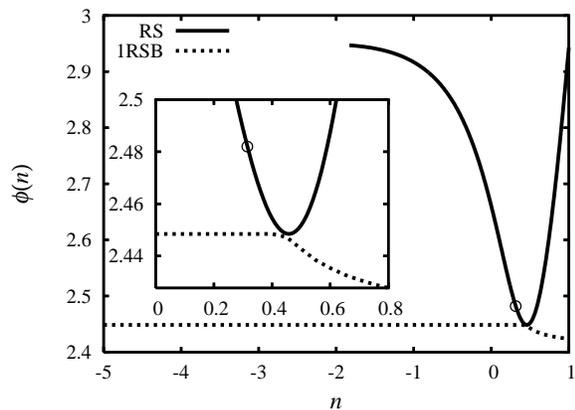}
\caption{$n$ dependence of the cumulant generating function $\phi$ for
 RS(solid line) and 1RSB(dashed line) solutions for $p=3$ and $\beta=3.0$. 
The full mark represents the AT instability point $n_{\rm AT}$. The
 inset shows an enlarged view around the minimum of the RS solutions. }
\label{negn}
\end{figure}

Similarly, we can perform the perturbation with respect to $\tau
 = \frac{T_c-T}{T_c}$. In fact, when we 
 substitute $\tau$ for $n$ in eqs.~(\ref{eqn:perturbation}), we
 can follow the above argument and conclude that $q_0(\tau,n)=\hat
 q_0(\tau,n)=0$ again. This implies that $\phi(\tau,n)=\phi(\tau,0)$ for any
 value of $n<0$ and in the vicinity of $\tau=0$, which is consistent
 with the results shown in ref.~\citen{DFM}.  
In particular, this means that $\phi(n)$ for $n<0$ is constant if $\tau$ is sufficiently
 small and  
a discontinuous phase transition for $q_0$ does not occur. To confirm
 this result, we examine the 1RSB solution for $p=3$ and $\beta=3.0$
 numerically by the steepest 
 descent method for the saddle point equations with respect to the order
 parameters. Figure\ \ref{negn} shows that it is certainly a constant
 function at least $-5<n<0$. 

 \section{Connection between RS and 1RSB solutions}
\label{sec:connection}
We have evaluated the cumulant generating function perturbatively under
the 1RSB ansatz in the previous section. This perturbative analysis would break down at a certain value
 of $n$ as $n$ increases from 0. To discuss a breaking point, 
 we approach it from the RS solution, which is the correct saddle point
 for $n$ larger than 1 at least.

The correct cumulant generating function $\phi$ generally has two
analytic properties, monotonicity of $\phi$ and convexity of $n\phi$, which can be
proven using H\"{o}lder's inequality as shown in Appendix~A. 
When one claims that an RS solution is valid at some value of $n$, the above
two properties have to be maintained in addition to the AT
stability.\cite{K} 
Therefore, we assume that the RS solution is valid if and only if  
these three properties hold. 

\begin{figure}
\hspace{-15pt}\includegraphics[width=8cm]{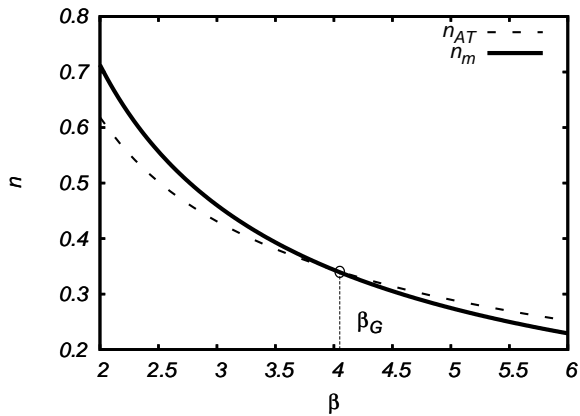}
\caption{$n_{\mathrm{AT}}(\beta)$ in dashed line and $n_m(\beta)$ in solid line for $p=3$. 
$\beta_{\rm G}$ indicates the inverse Gardner temperature. }
\label{m-AT}
\end{figure}

We examine whether the RS solution obtained using eq.~(\ref{phiRS}) satisfies these
properties. Let us first see the monotonicity.  
As shown in Fig.~\ref{negn}, it is clear that 
the monotonicity of the RS solution breaks down at some small positive value
of $n$, which we call $n_m(\beta)$.  
 This point is determined using the equation
 \begin{equation}
 \frac{{\rm d} \phi_0}{{\rm d} n} = \frac{\partial \phi_0}{\partial n} +
 \frac{\partial \phi_0}{\partial q}\frac{{\rm d} q}{{\rm d} n} + \frac{\partial
 \phi_0}{\partial \hat q}\frac{{\rm d}\hat q}{{\rm d} n} = \frac{\partial
 \phi_0}{\partial n} =0.
\label{eqn:mono}
 \end{equation}
The breaking point for other conditions is also determined using the
equations
\begin{equation}
\frac{{\rm d}^2 \phi_0}{{\rm d} n^2} = 0
\end{equation}
for the convexity and 
\begin{equation}
1=\frac{p(p-1)}{2}\beta^2q^{p-2}\frac{\int{\rm D} u\cosh^{n-4}\sqrt{\hat q}u}{\int{\rm D} u\cosh^n\sqrt{\hat q}u}
\label{nat}
\end{equation}
for the AT stability. We define $n_{\mathrm{AT}}(\beta)$ as the marginal solution
of the AT stability condition.
It is shown later that the convexity is not  broken when we set
parameters in the 1RSB stable regime.
Therefore, we consider which of the breaking points $n_m$ and $n_{\rm
AT}$ is relevant. As an example, we show $n_m$ and $n_{\rm AT}$ for
$p=3$ as a
function of the inverse temperature in Fig.~\ref{m-AT}. It is found that
the two lines intersect with each other at an inverse temperature, which
is denoted by $\beta_G$. 
We show later that this temperature $\beta_G$ is the Gardner
temperature. 
In the temperature region $\beta<\beta_G$, the monotonicity is first
broken 
 as $n$ decreases. 
Thus, we conclude that $\phi(n) = \phi_0(n)$ for $n> n_m$ in the 1RSB stable temperature
regime from the numerical evidence and the assumption mentioned
above. 

In other words, The above argument is that we take the RS solution to
 as small $n$ as possible, while the 1RSB solution should be valid in
 the vicinity of $n=0$. Then, we determine the cumulant generating
 function in the rest of the interval of $n$. It is, however,  easily
 obtained by the following proposition that holds in general: 
\newtheorem{prop}{Proposition}
 \begin{prop}
\label{prop:1}
 If $\phi'(n_0) = 0$ for a value of $n_0 > 0$, then $\phi(n) = \phi(0)$
  for $0<n<n_0$. 
\end{prop}
\noindent 
The proof is given 
in Appendix~\ref{sec:A}. 
Because we have $\phi_0'(n_m)=0$ in eq.~(\ref{eqn:mono}) by definition,
this proposition yields $\phi(n) =
 \phi_0(0)$ for $0<n<n_m$. It should be noted that this fact is derived
 without the 1RSB solution or the perturbation but with the plausible
 assumption.

Although the perturbation might not be necessarily required,  we use it to
 understand intuitively what is happening to the 1RSB solution.
From the perturbational argument, the order parameters turn out to be
 constant with respect to $n$. When we plot $q(x,n)$ as a function of $x$ with the
 value of $n$ changing, the shape of $q(x,n)$ is independent of $n$ as
 shown in Figs.~\ref{qxn1} and \ref{qxn2}. Because the lower limit of the domain of $q(x)$
 is $n$, $q(x)$ takes only one value  $q_1$ when $n=m$. This is how the
 1RSB solution connects to the RS one as $n$ increases from 0.   

\def\thefigure{3(a)}
\begin{figure}
\includegraphics[width=8cm]{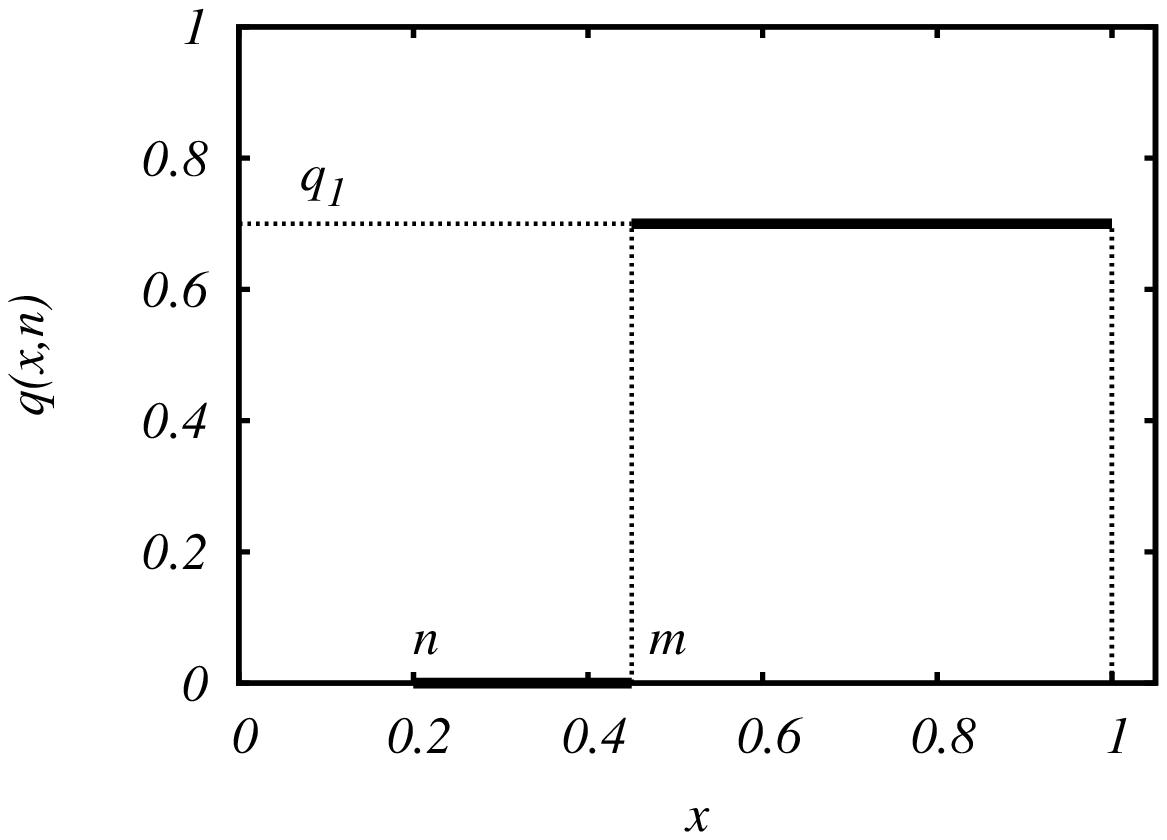}
\caption{}
\label{qxn1}
\end{figure}
\def\thefigure{3(b)}
\begin{figure}
\includegraphics[width=8cm]{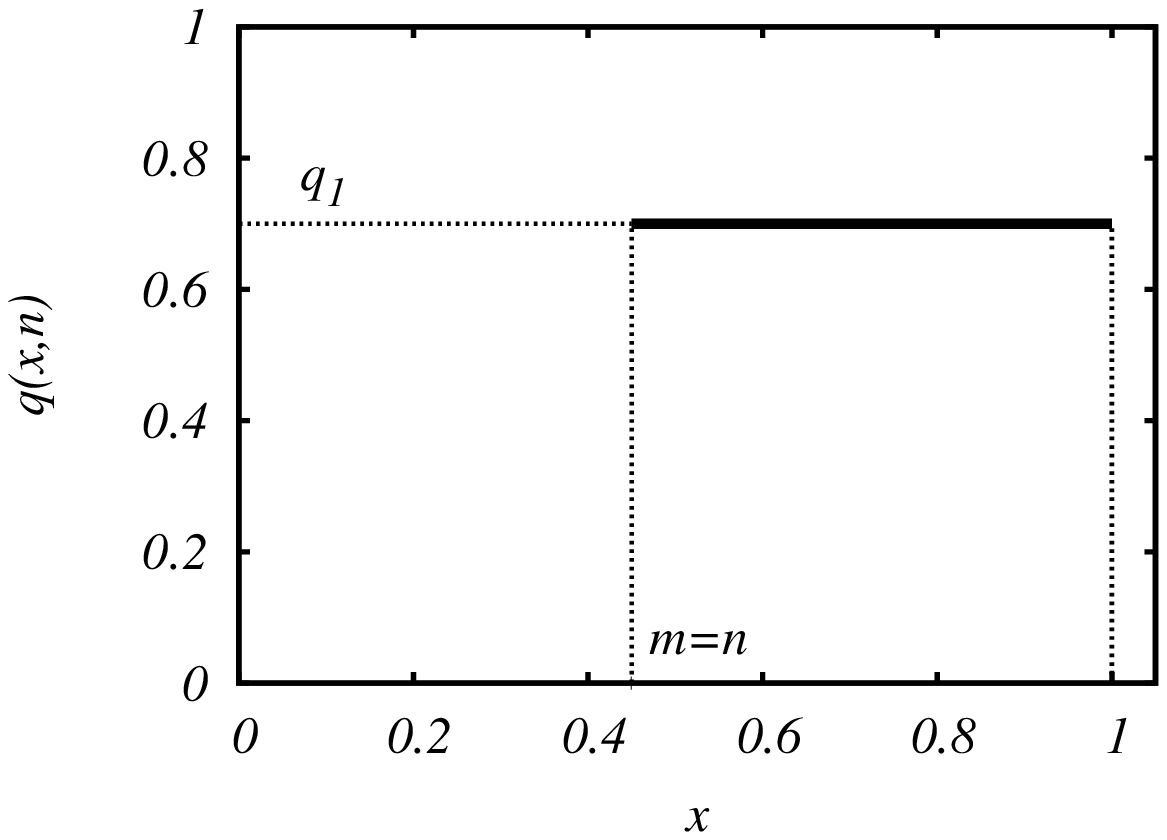}
\caption{Schematics of $q(x,n)$ for $n<m$\ (a) and $n=m$\ (b). }
\label{qxn2}
\end{figure}

We have discussed two continuations between the RS and 1RSB
solutions. One is from the RS solution to the 1RSB one with $n$
decreasing, which indicates that the connection point is $n_m$. The
other is from the 1RSB to the RS solutions with $n$ increasing from 0,
which indicates that it is $m$. Thus, this strongly suggests that $n_m =
m$. 

Let us compare the RS solution of eq.~(\ref{phiRS}) to the 1RSB
 solution with $q_0=0$ of eq.~(\ref{reduct1RSB}).  
They have the same expression by identifying $n$ with $m$, and $q,\hat
q$ with $q_1,\hat q_1$.  By definition, $n_m$ and $m$ give a minimal
value of $\phi_0$ and $\phi_1$, respectively.  Because
the other order parameters are also determined by minimizing respective
functions, we conclude that the RS solution at $n=n_m$ coincides with
the 1RSB solution including the order parameters. 
This yields certainly $n_m = m$. 

We now turn to the convexity of $n\phi_0$, which is written as
 \begin{eqnarray}
&\hspace{-150pt}\frac{{\rm d}^2\phi_0}{{\rm d} n^2}(n) =\nonumber\\
&\hspace{-15pt}\left(1,\frac{{\rm d} q}{{\rm d} n},\frac{{\rm d}\hat q}{{\rm d} n}\right)
{
\left(
\begin{array}{ccc}
\frac{\partial^2 \phi_0}{\partial n^2} &\frac{\partial^2 \phi_0}{\partial n\partial q} & \frac{\partial^2 \phi_0}{\partial n\partial \hat q}\\
\frac{\partial^2 \phi_0}{\partial n\partial q} &\frac{\partial^2 \phi_0}{\partial q^2} & \frac{\partial^2 \phi_0}{\partial q\partial \hat q}\\
\frac{\partial^2 \phi_0}{\partial n\partial \hat q} &\frac{\partial^2 \phi_0}{\partial q \partial \hat q} & \frac{\partial^2 \phi_0}{\partial \hat q^2}
\end{array}
\right)
\left(
\begin{array}{c}
1\\
\frac{{\rm d} q}{{\rm d} n}\\
\frac{{\rm d}\hat q}{{\rm d} n}
\end{array}
\right)
} \nonumber\\
&\hspace{150pt}\geq 0.
 \end{eqnarray}
 From the correspondence mentioned above, the matrix in this inequality
 is the same as the Hessian of the 1RSB solution under
 the condition $q_0=\hat q_0=0$ at $n=0$. The matrix must be positive definite because of the  stability of 
 the 1RSB solution in the parameter region $\beta_c<\beta<\beta_G$. 
Then,  we do  not have to consider the convexity of the RS solution. 

Let us next show that the temperature at which the lines $n_m(\beta)$ and
 $n_{\mathrm{AT}}(\beta)$ cross is the Gardner temperature. 
We have defined $n_{\mathrm{AT}}$ as the solution of eq.~(\ref{nat}) and 
 have shown $n_m = m$.  Substituting $n_m = m$ for $n$ in
 eq.~(\ref{nat}), we obtain that the equation is identical to 
the AT stability condition of the 1RSB
 solution\cite{G}. Therefore, it is shown that the AT stability condition
 of the RS solution at $(\beta, n_m(\beta))$ is equivalent to that of
 the 1RSB solution at $(\beta, 0)$. This implies that the Gardner
 temperature is determined by the crossing temperature of $n_m(\beta)$
 and $n_{\mathrm{AT}}(\beta)$. 
We note that the instability of the 1RSB solution is given by the
 analytic properties of the RS solution at a finite value of $n$. 

 \section{Calculation of Rate Function}
\label{sec:rate}

We have the cumulant generating function $\phi(n)$ for any value of $n$ as 
  \begin{equation}
  \phi(n) =\left\{
\begin{array}{ll}
\phi_0(n) &  (n>n_m),\\ 
\phi_1(n) =\phi_0(n_m) & (n<n_m), \def\nonumber{\global\@eqnswtrue}
\end{array}
\right. 
		  \label{phi}
  \end{equation}
although an explicit formula is not given.  As an example, we show in
		  Fig. \ref{pbodyCGF} the cumulant generating function
		  evaluated numerically for $p=3$. 
The function for the
		  random energy model, corresponding to $p\to\infty$ in
		  our model, is given rigorously as\cite{T}
  \begin{eqnarray*}
  \lim_{p\to\infty}\phi(n) =\left\{
\begin{array}{ll}
\phi_0(n)=\frac{\beta^2}{4}n + \frac{\log 2}{n}\hspace{12pt} &\hspace{-20pt} (n>n_m),\\ 
\phi_1(n)=\beta\sqrt{\log 2}, \def\nonumber{\global\@eqnswtrue} &\hspace{-20pt} (n<n_m)
\end{array}
\right. 
  \end{eqnarray*}
  for $\beta>\beta_c$.
Thus, it is confirmed that $\phi(n)$ is constant for $n<0$ from the three
  independent results, perturbation with respect to $\tau$, numerical calculations, and the exact solution for $p\to\infty$.

\def\thefigure{4}
\begin{figure}
\includegraphics[width=8cm]{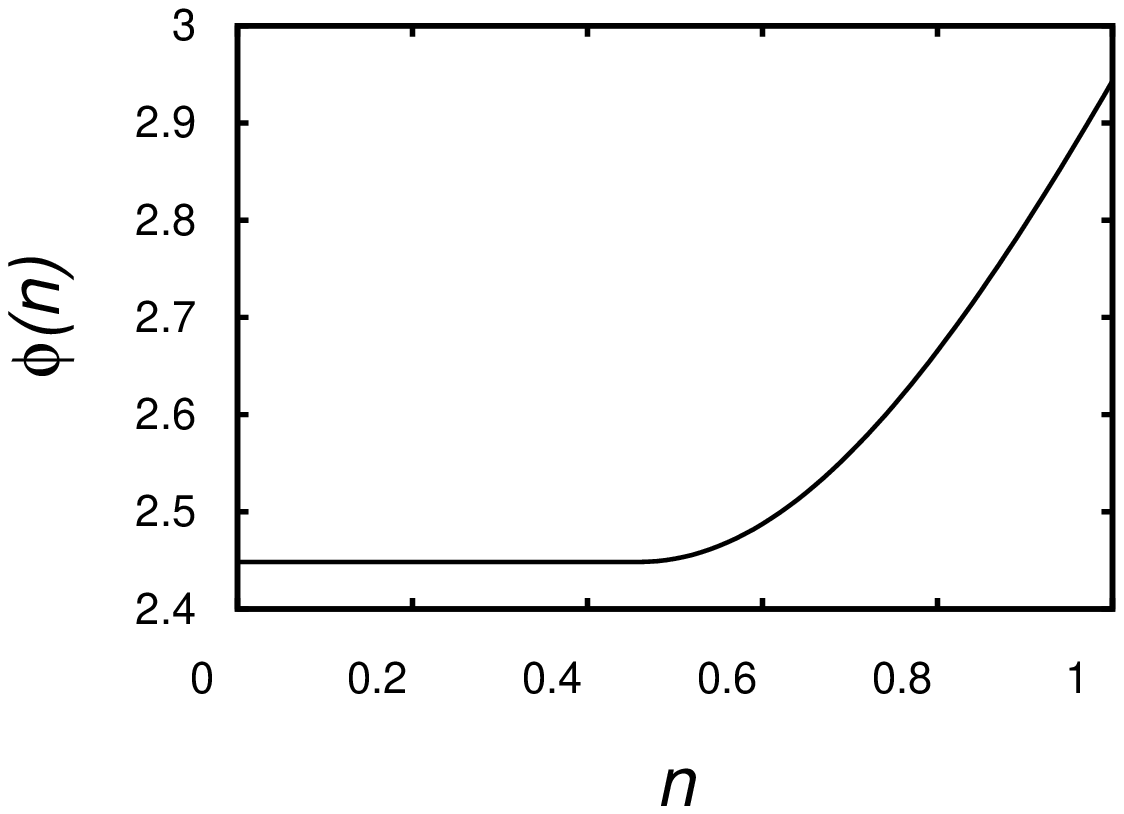}
\caption{Cumulant generating function for $p = 3$ and $\beta=3.0$.}
\label{pbodyCGF}
\end{figure}

By means of eqs. (\ref{Legendre}) and (\ref{phi}), the rate function is
derived in principle.
Let us first evaluate the rate function for a small fluctuation around the
thermodynamic value $f_0 =
\lim_{N\to\infty}\frac{1}{N}\log Z$. 
Noting that $\phi(n_m) = \phi(0) = f_0$ and $\phi'(n_m) = 0$, the
equation $f = \frac{{\rm d} (n\phi(n))}{{\rm d} n}$ at $n=n_m+dn$
leads to 
\begin{eqnarray*}
f_0 + {\rm d} f &=& \phi(n_m+{\rm d} n) + (n_m+{\rm d} n)\phi'(n_m+{\rm d} n) \\
		   &=& f_0 + n_m\phi''(n_m){\rm d} n  + O({\rm d} n^2), 
\end{eqnarray*}
where ${\rm d} f$ is a small and positive variable. 
Eventually, we  have 
\begin{equation}
{\rm d} n = \frac{1}{n_m\phi''(n_m)} {\rm d} f, 
\end{equation}
and the rate function around $f_0$ is obtained as
\begin{eqnarray*}
\Sigma(f_0 + {\rm d} f) \hspace{-10pt}&=&\hspace{-10pt} (n_m+{\rm d} n)(f_0 + {\rm d} f - \phi(n_m + {\rm d} n))\\
				\hspace{-10pt}&=&\hspace{-10pt} n_m {\rm d} f. \def\nonumber{\global\@eqnswtrue}
				\label{rf}
\end{eqnarray*}
Thus, we conclude that the rate function has a linear term 
with respect to the fluctuation around the 
 thermodynamic value and its coefficient is $n_m$. 
This is in contrast with the SK model with $p=2$, where the leading term
 of $\Sigma(f_0+{\rm d} f)$ is anomalously $({\rm d} f)^{(6/5)}$\cite{PR} although other
 possibilities have also been discussed. 
It would be interesting to see that the finite value of the replica
 number $n$, which originates from RM, has a physical significance before 
taking the limit of $n$ to $0$.

The linear dependence of the rate function reminds us of
a structural similarity to the thermodynamics. 
Rewriting the definition of $\phi(n)$ and $\Sigma(f)$,  we have 
\begin{eqnarray*}
-\frac{N}{\beta}\phi(n) &\hspace{-10pt}\approx&\hspace{-10pt} -\frac{1}{\beta n} \log
\left[\exp\left\{-(\beta n)\left(-\frac{1}{\beta}\log Z\right)\right\}\right],  \\
-N \Sigma (f) &\hspace{-10pt}\approx&\hspace{-10pt} \log \mathrm{Pr}\left(\frac{1}{N}\log Z \approx
f\right). 
\end{eqnarray*}
They can be regarded as the ``free energy'' and the ``entropy'',
respectively,  when we identify the true free energy $-\frac{1}{\beta}\log Z$ as the
 Hamiltonian. 
Then, $\beta n$ is an effective ``temperature'' in this sense and 
eq. (\ref{rf}) is interpreted as ``$T {\rm d} S = {\rm d} E$'', because we
use $f$ as $\log Z$ and $\Sigma(f_0) = 0$. 
Further comparison to thermodynamics  is discussed in Appendix B.

Let us complete the calculation of $\Sigma(f)$. When $f< f_0$, $\Sigma(f) = \infty$ because
 $\phi(n) = \phi(0)$ for $n<0$. When $f\gg 1$, $\Sigma(f) \approx \frac{f^2}{4\alpha}$ with $\alpha$ being some positive constant. This is because
 $\phi(n)$ is approximately proportional to $n$ with the coefficient
 $\alpha$ when $n$ is sufficiently
 large. 
This property is derived from these inequalities\cite{vHP}:
\begin{eqnarray*}
[Z^n] &\hspace{-10pt}\geq&\hspace{-10pt} [\exp(-n\beta H_1)]  = \exp\left(\frac{\beta^2}{4}Nn^2\right) \def\nonumber{\global\@eqnswtrue}
\end{eqnarray*}
where $H_1$ is an energy value for a given configuration of $\{S_i\}$  
and for $n\in {\bf N}$
\begin{eqnarray*}
\hspace{-10pt}[Z^n] &\hspace{-10pt}=&\hspace{-10pt} \mathrm {Tr} [\exp(-\sum_{a=1}^n\beta H^{(a)})] \\
	&\hspace{-10pt}=&\hspace{-10pt} \mathrm{Tr} \exp\left\{\frac{\beta^2p!}{4N^{p-1}}\sum_{i_1<\cdots <i_p}\left(\sum_{a=1}^nS_{i_1}^{(a)}\cdots S_{i_p}^{(a)}\right)^2\right\}\\
	&\hspace{-10pt}\leq &\hspace{-10pt} 2^{Nn} \exp\left(\frac{\beta^2}{4}Nn^2\right). \def\nonumber{\global\@eqnswtrue}
\end{eqnarray*}
Thus, for all $n'\in {\bf R}$, $\phi(n') \leq \phi(n+1) \leq\frac{\beta^2}{4}(n+1) + \log 2$, where $n$ is 
the largest integer which does not exceed $n'$. 
These imply that $\alpha = \frac{\beta^2}{4}$. 

To fix the normalization factor ${\cal N}$, we introduce a
characteristic value $F$ of the free energy at which the expression of
$\Sigma(f)$ changes. 
For sufficiently large $N$, the probability distribution is expressed as
(\ref{pdf}) with $\Sigma(f)$. Then,  the normalization condition can be
approximately written as 
\begin{eqnarray*}
1 &\hspace{-10pt}=&\hspace{-10pt} \int_{f_0}^F {\rm d} f {\cal N} {\rm e}^{-N n_m (f-f_0)} + \int_F^\infty {\rm d} f
{\cal N} {\rm e}^{-N\alpha f^2}  \\
  &\hspace{-10pt}=&\hspace{-10pt} \int_{0}^{N(F-f_0)} \frac{{\rm d} f}{N} {\cal N} {\rm e}^{-n_m f} + \int_{\sqrt{N}F}^\infty \frac{{\rm d} f}{\sqrt{N}} 
{\cal N} {\rm e}^{-\alpha f^2}.  
\end{eqnarray*}
Because the second term is a Gaussian integral whose integral range does
not include its peak, it vanishes exponentially as $N$ increases. Then,
we find that the leading term of $\cal N$ is $Nn_m$. 
By using this normalization factor, the mean of the fluctuation $f_0 + {\rm d}
f$ is evaluated as 
\begin{equation}
[{\rm d} f] = \frac{1}{Nn_m}.
\end{equation}
Thus it decays as $N^{-1}$ and this should be confirmed numerically.
\section{Summary and Conclusions}
\label{sec:summary}
In this paper, we have calculated the cumulant generating function and
 the rate function of $p$-body  
 Sherrington-Kirkpatrick model in the 1RSB temperature regime and argued the probabilistic property of 
the free energy $\log Z$.

We have evaluated the cumulant generating function by using RM with the
 replica number $n$ finite from the two viewpoints.
The perturbation  analysis based on the 1RSB solution near $n=0$
 indicates that the free energy and the order parameters are constant as a
 function of $n$ in the vicinity of $n=0$. 
From the assumption that the RS solution is valid for $n>n_m$ when
 $\beta<\beta_G$, it follows that if we
 can take the RS solution down to $n=n_m$, the cumulant generating 
function for $0<n<n_m$ is determined uniquely as a constant function.

Performing the Legendre transformation, we have evaluated the rate
function, whose leading term near the thermodynamic value $f_0$ have been found
to be $\Sigma(f_0+{\rm d} f) = n_m {\rm d} f$ for $0<{\rm d} f \ll 1$. This form is
similar to the conventional thermodynamics if we regard the free energy as the ``Hamiltonian''.

From these arguments, we have considered that the monotonicity breaking point $n_m$
is more relevant than the AT stability  
breaking point $n_{\mathrm{AT}}$ when the temperature is in the 1RSB regime. It is
because the crossing point of $n_m$ and $n_{\mathrm{AT}}$ as a function of
temperature is identical with the Gardner temperature at least in the
model discussed. This criterion for determining the Gardner temperature
is formally equivalent to that shown in ref.~\citen{KZ}, where a finite-temperature
phase diagram of a coloring problem on a finite connectivity graph is
discussed. We suppose that the monotonicity breaking explains the reason
why the breaking parameter $m$ is determined ``thermodynamically''.

Furthermore, this is one of the possible mechanisms of how an RS solution
breaks to a 1RSB solution. Note that it is not based on the AT stability
breaking, but on the monotonicity breaking. 
The latter case yields only the 1RSB solution in principle. 
This scenario might explain why the RS solution breaks down only one
step in many other models. In other words, the 1RSB solution could be
derived from the RS solution fixed as a monotonically increasing
function. Thus, using this scheme, one may calculate the ``1RSB'' free
energy even if the standard 1RSB Parisi matrix is hardly constructed, which
is the case in such models as finite-connectivity systems.


\acknowledgements
We would like to thank Y. Kabashima for the helpful comments and
discussions. This work was supported by a Grant-in-Aid for Scientific
Research on the Priority Area ``Deepening and Expansion of  Statistical
Mechanical Informatics'' from the Ministry of Education, Culture, Sports,
Science and Technology. 

\appendix
\section{Proof of Proposition 1}
\label{sec:A}
In the Appendix, we prove the proposition \ref{prop:1} mentioned in
\S\ref{sec:connection}. 
The proof relies on the following lemmas concerning the general
properties of the cumulant generating function: 
\begin{enumerate}
 \item (monotonicity) \textit{$\phi(n)$ is a monotonically increasing function in a broad
       sense.}
 \item (convexity) \textit{ $n\phi(n)$ is a convex function in a broad sense.}
\end{enumerate}

Their proofs are as follows: From the H\"older's inequality,  we
have  for $n<m$ 
\begin{equation}
[Z^n] = [Z^{m\frac{n}{m}}] < [Z^m]^{\frac{n}{m}}.
\end{equation}
From dividing the logarithm of both sides by $nN$, it follows that 
$\phi_N(n) <\phi_N(m)$. In the limit $N\rightarrow\infty$, 
the monotonicity is proven.

From the H\"older's inequality again, for $s<t$ and $0<\alpha<1$, we
have  
\begin{equation}
[Z^{\alpha s + (1-\alpha) t}] < [Z^s]^{\alpha} [Z^t]^{1-\alpha}.  
\end{equation}
In a similar manner, this leads to 
\[
(\alpha s + (1-\alpha) t)\phi_N(\alpha s + (1-\alpha) t) <\alpha s\phi_N(s) + (1-\alpha)t\phi_N(t).
\]
In the limit $N\rightarrow\infty$,  this proves the convexity.

From the monotonicity, we have 
\begin{equation}
\phi(n) \leq \phi(n_0)
\label{monotone}
\end{equation}
for  $n<n_0$, while  from the convexity, we have 
\[
n\phi(n) \geq (n_0\phi'(n_0) + \phi(n_0))(n-n_0) + n_0\phi(n_0), 
\]
where the right-hand side represents the tangent line at $n=n_0$. Because
$\phi'(n_0) = 0$, this inequality is reduced to $n\phi(n) \geq
n\phi(n_0)$. It turns out that for $n>0$, 
\begin{equation}
\phi(n) \geq \phi(n_0). 
\label{conv}
\end{equation}
The conditions (\ref{monotone}) and (\ref{conv}) are satisfied
simultaneously only when $\phi(n) = \phi(n_0)$. This completes the proof
of the proposition. 

\section{Entropy Interpretation}

Here, we comment on the non-negativity condition of the rate function,
$\Sigma(f)\geq 0$, which is derived in ref.~\citen{IHOK} from the fact
that the probability distribution function has to be normalizable.  
From this condition, the cumulant generating function $\phi(n)$ is
restricted as  
\[
\frac{{\rm d}(n\phi(n))}{{\rm d} n} \geq \phi(n), 
\]
which is reduced to 
\[
n\phi'(n)\geq 0.
\]
Thus, we see that the non-negativity of $\Sigma(f)$ corresponds to 
the monotonicity condition of $\phi(n)$ for $n\geq 0$.

This condition is a kind of ``entropy crisis'', because $\Sigma$ can be
 interpreted as entropy induced by the randomness. We
 also consider the condition imposed from the real entropy
 crisis $S = \beta^2\frac{{\rm d}}{{\rm d} \beta} \left(-\frac{1}{\beta}\log Z\right)\geq 0$
 , which is \[
\frac{\partial}{\partial \beta}\psi(\beta,\gamma)\geq 0,
\]
 where 
\begin{eqnarray*}
\psi(\beta,\gamma) &:=& -\frac{1}{\gamma}\log [\exp(-\gamma F(\beta))]\\
					&=& -\frac{1}{\beta}\phi(\gamma/\beta).
\end{eqnarray*}
 Note that the sign of $\frac{\partial \phi}{\partial \beta}$ is not determined. In this sense, this paraphrase suggests that $\gamma=\beta n$ is a more natural parameter than 
 $n$.

\end{document}